# Measuring Thermal Profiles in High Explosives using Neural Networks

J. Greenhall[1,a], D. K. Zerkle[1], E. S. Davis[1], R. Broilo[1], and C. Pantea[1]

[1]Los Alamos National Laboratory, Los Alamos, NM, 87545

[a]Corresponding author: John Greenhall, jgreenhall@lanl.gov

**Abstract**

We present a new method for calculating the temperature profile in high explosive (HE) material using a Convolutional Neural Network (CNN). To train/test the CNN, we have developed a hybrid experiment/simulation method for collecting acoustic and temperature data. We experimentally heat cylindrical containers of HE material until detonation/deflagration, where we continuously measure the acoustic bursts through the HE using multiple acoustic transducers lined around the exterior container circumference. However, measuring the temperature profile in the HE in experiment would require inserting a high number of thermal probes, which would disrupt the heating process. Thus, we use two thermal probes, one at the HE center and one at the wall. We then use finite element simulation of the heating process to calculate the temperature distribution, and correct the simulated temperatures based on the experimental center and wall temperatures. We calculate temperature errors on the order of 15°C, which is approximately 12% of the range of temperatures in the experiment. We also investigate how the algorithm accuracy is affected by the number of acoustic receivers used to collect each measurement and the resolution of the temperature prediction. This work provides a means of assessing the safety status of HE material, which cannot be achieved using existing temperature measurement methods. Additionally, it has implications for range of other applications where internal temperature profile measurements would provide critical information. These applications include detecting chemical reactions, observing thermodynamic processes like combustion, monitoring metal or plastic casting, determining the energy density in thermal storage capsules, and identifying abnormal battery operation.

## 1. Introduction

Noninvasive measurement of internal temperature distribution is critical to a range of applications, including detecting chemical reactions, observing thermodynamic processes like combustion, monitoring metal or plastic casting, determining the energy density in thermal storage capsules, identifying abnormal battery operation, and assessing the safety status of high explosives (HE). Currently, there are no good noninvasive techniques for measuring temperature distribution in a sealed container.

Classical thermometry techniques are limited to measuring the outside temperature of the container or require puncturing the container, which can interfere with the process being monitored and pose a safety hazard. Additionally, these techniques are typically limited in the number of internal locations where temperature can be measured, and the embedded instruments can interfere with the process being monitored. Alternatively, acoustic techniques have been developed to enable measuring temperature distributions at an arbitrary number of internal points without interfering with the physical process.[1–5] These techniques are based on acoustic Time-of-Flight (ToF) measurements using an array of acoustic transducers. One at a time, each transducer transmits an acoustic burst, which then propagates through the material to the receivers. The time required for the acoustic bursts to travel between each transmitter/receiver pair is dependent on the sound speed throughout the material, which is dependent on the temperature distribution. The temperature is measured using either a 2-step or 3-step process. In the 2-step process, the sound speed distribution is calculated directly from the measured waveforms using techniques such as full-waveform inversion[6–8] or a convolutional encoder-decoder network,[9] and then temperature is determined from sound speed using an empirical model for the given material. In the 3-step process, the ToF is measured from the waveforms, the sound speed distribution is determined using reverse-time migration.[10–12] and then the temperature is calculated from an empirical model. However, demonstration of the existing acoustic methods is limited to measuring temperature in single-phase (gas, liquid, or solid) materials, and the techniques require the transducers to be in direct contact with the material.

In contrast, many applications require temperature distribution measurements in other materials, and they require a noninvasive measurement, i.e. transducers must measure through the container walls. In this case, some of the acoustic burst energy propagates through the internal material as a bulk wave, while the remaining energy travels through the container walls as guided waves. As a result, the guided waves interfere with the bulk waves, which inhibits implementing waveform inversion or reverse-time migration. When measuring highly attenuating materials, lower acoustic frequencies are required, which increases the burst durations and further increases the overlap between different arrivals. In previous work, bulk wave arrivals were isolated by using cross-correlation with broad-band chirps [13] or using a Convolutional Neural Network (CNN).[14] However, to measure sound speed, and, thus temperature, these techniques still require the use of reverse-time migration, which can be highly sensitive to the initial sound speed estimate and to error in the estimated arrival time.

To overcome the limitations of existing temperature measurement techniques, we present a novel technique based on time-domain acoustic measurements processed via CNN. In contrast with traditional temperature sensors and existing acoustic methods, our technique enables measuring the temperature profile through a material, it is noninvasive, and it works for challenging, highly attenuating materials such as HE. To train and test the technique, we utilize a novel mixture of experimental and simulated data to provide acoustic and temperature profile measurements, respectively. We conduct experiments/simulations of a cylindrical container filled with HE (pentolite 50/50) as it is heated to the point of detonation or deflagration to provide a variety of thermal profiles. This technique enables measuring real-time temperature profiles in a material noninvasively, which is not possible using existing techniques. In addition to monitoring the safety status of HE, this technique could be invaluable for a wide range of other applications including, assessing capacity in thermal storage systems, quantifying performance in molten salt reactors, measuring chemical kinetics, and monitoring material composition in various industrial processes, to name a few.

## 2. Methods
### a. Experimental acoustic and thermocouple measurements

The goal of this work is to use a CNN to measure the temperature profile within the HE based on the acoustic bursts transmitted between an acoustic transmitter (Tx) and one or more receivers (Rx). To enable CNN training and testing, we will acquire hybrid experimental/simulation data. Figure 1 shows the experimental data collection process. A cylindrical container (Al-6061) with dimensions, 144 mm inner diameter ($2R$), 6.4 mm thickness, 200 mm height is equipped with 16 piezoelectric transducers (STEMINC SMD07T05R411), evenly spaced around the container circumference, and two thermocouples are inserted into the HE at the wall ($r = R$) and center ($r = 0$) at approximately the same height as the acoustic transducers (Figure 1(a)). Over the course of an experiment, heaters placed at the bottom of the container gradually heat the HE until it detonates or deflagrates. During each experiment, we collect a set of acoustic (Figure 1(b)-(c)) and thermocouple measurements (Figure 1(d)) at 2 min intervals. Due to the high acoustic attenuation within the HE, must select a relatively low excitation frequency.[14] We utilize a Gaussian burst with 10 $V_{pp}$ amplitude, 350 kHz center frequency, 150 kHz standard deviation. Figure 1(b) shows a cross-section of the acoustic waves propagating through the HE and container from one transmitter (Tx) to the remaining 15 receivers (Rx). Figure 1(c) shows an example acoustic measurement, which consists of 15 waveforms, transmitted from one Rx and received from the remaining Rx, with lines indicating the theoretical arrival times for the first bulk (red) and guided waves (green). At each time step in the experiment, we repeat this acoustic measurement, using each of the Tx, one at a time.

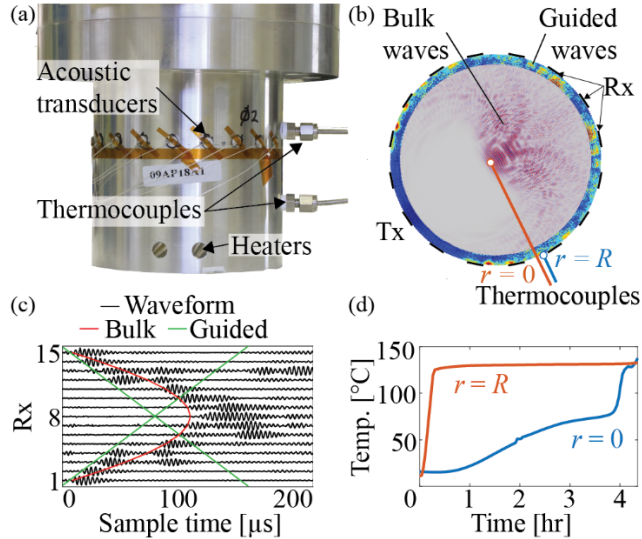

Figure 1: Experimental data collection. (a) A container filled with HE is heated from below. (b) A cross-section shows the acoustic bulk and guided waves transmitted/received between 16 acoustic transducers. (c) Example acoustic measurement from one Tx to 15 Rx. (d) Two thermocouples measure HE temperature at the wall and center of the container over the course of the experiment.

### b. Simulated temperature profiles in HE

The thermocouple data provides temperature information at two locations $r = 0$ and $R$ (Figure 1(d)), but measuring the temperature profile with a useful amount of radial resolution would require a significant number of thermocouples that would interfere with the HE heating process. Thus, to acquire temperature profiles, we employ axisymmetric numerical simulations in COMSOL, based on an existing HE modeling methodologies which account for the heat transfer, phase change, species transfer, and natural convection within the HE.[15,16] Figure 2 shows the numerical simulation setup for the axisymmetric HE container. We utilize the built-in heat transfer module to simulate the HE and container temperatures as they are heated from approximately 20 °C by the heater, which is represented by ramping up and then holding the temperature at the heater boundary to 180 °C. Pentolite 50/50 consists of TNT (50%), which has a melting temperature of approximately 80 °C and PETN (50%), which melts at approximately 140 °C. Thus, when the temperature 80°C<T<140°C the TNT melts, and the embedded PETN particles begin to sink. When the temperature exceeds 140°C, the PETN also melts, and the two species can diffuse into one another. This results in gradients in the material concentrations, which we represent using the species transport and laminar flow mixture model modules in COMSOL.

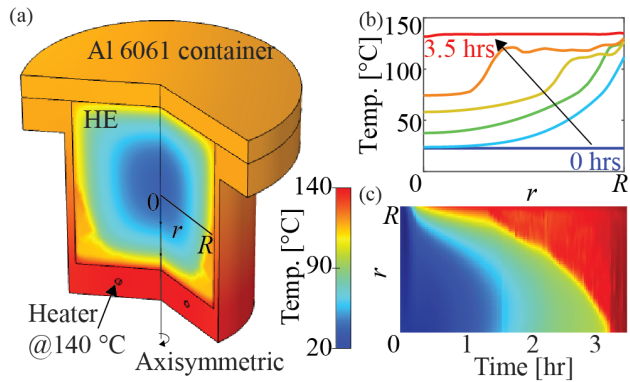

Figure 2: Simulated HE heating. (a) An axisymmetric finite element model used to compute the temperature distribution at the transducer cross-section as a function of radial position $r$. (b) Selected example radial temperature profiles at various experiment times. (c) Colorplot showing the temperature as a function of radial position $r$ and time over the course of the experiment.

After completing the simulation, we select a line from $r = 0$ to $r = R$, at the same height that the acoustic transducers are mounted. Figure 2(b) shows several radial temperature profiles at various steps throughout the experiment. Figure 2(c) shows a colorplot of the temperature as a function of radial position and time throughout the experiment.

### c. Machine learning with hybrid measurements

Prior to performing ML, we preprocess the experimental and simulated measurements as illustrated in Figure 3. Each acoustic measurement consists of an $N_t \times N_{Rx}$ array of waveforms, where $N_{Rx}$ is the number of waveforms used, and each waveform is a time series of length $N_t$. We investigate the effect of the number $N_{Rx}$ of Rx measurements used, where we select $N_{Rx}$ to be an odd number of Rx opposing Tx. We then reduce the noise amplitude and emphasize acoustic signals that are similar to the excitation $x_{ex}$ by cross-correlating the raw waveforms $X_w$ with the excitation signal to get $X_{cc} = X_w * x_e$, where * is the cross-correlation operator.[17] We then reduce the number of peaks and the range of feature scales within the experimental acoustic measurements, by computing the envelopes $X_e$

$$X_e = |H(X_{cc})(t)|, \quad (1)$$

where $H(\cdot)(t)$ denotes the Hilbert transform. The input signal $X$ to the CNN model is then created by normalizing the envelopes based on the standard deviation for each Rx

$$X = \frac{\sqrt{N_t}(X_e - \bar{X}_e)}{\sqrt{\sum(X_e - \bar{X}_e)^2}}, \quad (2)$$

where the bar $\bar{X}_e$ indicates the mean value of $X_e$ over time for a single Rx value.

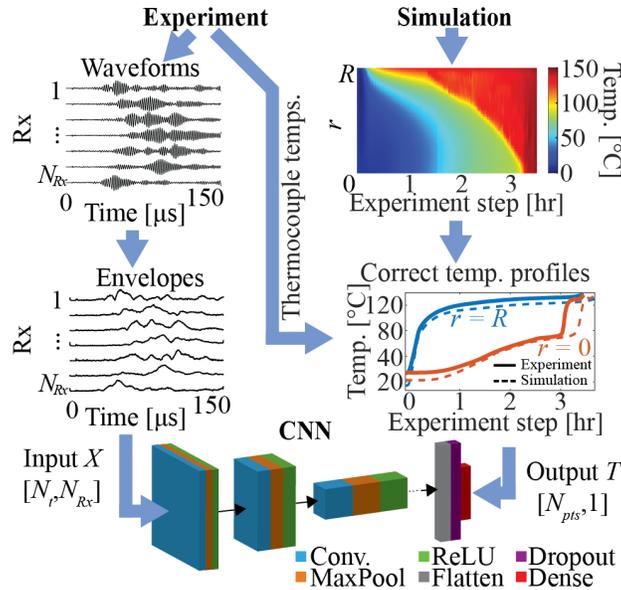

Figure 3: Preprocessing and machine learning steps for hybrid measurements from experiment and simulation.

To preprocess the temperature profiles, we need to correct for error between the temperatures from experiment and simulation. These are typically due to differences in HE material properties due to the casting process, inconsistent input power, non-axisymmetric components, defects, or physics in the experiment, or electrical noise in the heaters or thermocouples. Figure 3(right) shows an example of experimental (solid) and simulated (dashed) temperature profiles at the wall (blue) and HE center (orange). To account for differences, we correct the simulated temperature profiles based on the experimental thermocouple temperatures at the boundaries. We first calculate parameters to shift and scale the initial uncorrected simulation temperatures $T'(r,s)$ at position $r$ and experiment step $s$ to match the experimental temperatures at the container boundaries ($r = 0, R$). To simplify the formulae, we adopt a subscript $r$ notation, which indicates a term that is a variable of $r$ is being evaluated at a boundary, e.g. $T_r(s)$ where $r = 0$ or $R$ indicates the center or wall boundaries, respectively. The corrected simulated temperatures $T_r(s)$ at the boundaries can be calculated as

$$T_r(s) = a_r \cdot \{T'_r(c \cdot [s - d]) - b_r\}. \tag{3}$$

Here, temperature scale $a_r(s)$ and shift $b_r(s)$ and time scale $c(s)$ and shift $d(s)$, and temperature coefficients are linearly interpolated between the boundary values at $r = 0$ and $R$. We group the scale and shift coefficients at the boundaries into a single set $\theta = \{a_0, b_0, a_R, b_R, c, d\}$, and the optimal $\theta^*$ is computed by minimizing the mean-squared error over experiment steps $s$ between the corrected simulated temperatures and the experimental temperatures at the boundaries,

$$\theta^* = \underset{\theta}{\mathrm{argmin}} \sum_{r=0,R} \|T_r - \hat{T}_r\|^2. \tag{4}$$

Here, $\hat{T}_r$ is the experimental boundary temperatures. Finally, we will investigate the effect of the temperature resolution, i.e. the number $N_{pts}$ of radial points at which the CNN estimates the temperature. To train the CNN using different values of $N_{pts}$, we can simply interpolate the radial temperature profiles from COMSOL at $N_{pts}$ locations in $0 \leq r \leq R$.

As a result of the hybrid experimental/simulated data process, we obtain input data $X$ and output temperature profile data $T$ that can be used to train and test the CNN. Figure 3(bottom) shows the CNN architecture, which consists of a series of CNN blocks, each comprising of a 2D convolution (Conv) layer, a rectified linear unit (ReLU) activation layer, and a pooling (MaxPool) layer. At each Conv block, input data (think acoustic time-series data from multiple receivers) is convolved by a series of convolutional filters (see reference for detailed formulae[18]). The intent is for the filters to identify patterns within each acoustic signal and between neighboring signals and then transform the input signals to accentuate useful signal features and suppress signal noise. Next, the convolved data is passed to the ReLU layer, which introduces nonlinearities into the model that increases learning speed and performance.[19,20] The data passes through a MaxPool layer, which effectively returns a summary of the input data, which has been reduced in size so as to reduce the CNN model complexity and reduce the model sensitivity to slight shifts in the input data. We implement three CNN blocks, where the Conv layers consist of $8*2^{(l-1)}$ filters with dimension 16×2 for each layer $l = 1, 2$, and 3. By utilizing multiple CNN blocks in series, it is possible to reduce complex acoustic signals to one or more extracted features that represent the critical information conveyed by the acoustic signal. After the final CNN block, we then flatten the signal to a 1D array, apply a Dropout layers to ensure that the network does not rely too heavily on any one neuron during training, and then use a dense Output layer that applies a linear transformation between the flattened features and the estimated temperatures. This produces an $N_{pts}$×1 output $T$ representing temperatures at each of the radial points.

Because of small differences in the transducer geometry, material properties, positions, and adhesive layer dimensions and material properties, there are variations in the transfer functions between pairs of transducers. Our goal is to develop a temperature measurement method that is robust to these differences, as well as differences in the container and HE. Thus, we divide the data into sets, where each set consists of the measurements from a single transmitter to all receivers for all measurements in a single heating experiment. As a result, the combination of Tx/Rx transmission functions is unique between data sets. We then group the data sets randomly into 10 folds to test using k-folds cross-validation, wherein we train the model on all but one folds and test on the excluded fold, for each combination of training/testing folds.

## 3. Results

We perform the cross-validation procedure, training on all-but-one fold and estimating the temperature distributions on the remaining fold, for each combination of folds. Figure 4(a) shows some example radial temperature profiles from a single test set, i.e. the acoustic measurements from a single Tx over the course of a single heating experiment. Here, we have selected the results using $N_{Rx} = 3$ opposing receivers and a special resolution of $N_{pts} = 25$ radial points between $r = 0$-$R$. We plot the radial temperature distributions at several experiment steps $s$ as the HE was heated from approximately 20 °C ($s_0$, dark blue) until detonation/deflagration ($s_{max}$, red), where the dashed and solid lines indicate the temperature profiles estimated by the CNN and the "true" temperature profiles simulated in COMSOL. We observe small errors between the estimated and true temperature profiles, which are likely due to differences between the axisymmetric simulated temperatures and the experimental temperatures, as well as differences in the transducer-wall coupling between Tx-Rx data sets used in training versus testing. Despite the small errors, we observe that the CNN is able to closely estimate the temperature trends, i.e. where the temperature slope is steep/flat. This is an important finding because it indicates where the solid-liquid HE transition occurs, which provides critical information about the status of the HE.

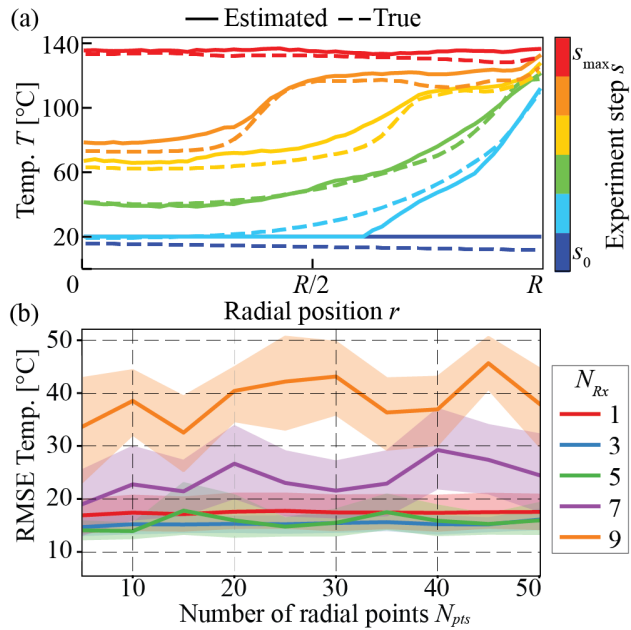

Figure 4: CNN testing results. (a) Example estimated (solid) and true (dashed) temperature profiles at several times throughout one experiment. (b) Mean temperature error vs number of radial points $N_{pts}$ at which temperature was estimated for different numbers of receivers $N_{Rx}$.

In addition to a qualitative comparison, evaluated the effect of the number of radial points $N_{pts}$ and number of receivers $N_{Rx}$ on the Root-Mean Squared Error (RMSE) between the true temperatures and those estimated by the CNN. For each we tested combinations of ($N_{pts}$, $N_{Rx}$) values in the ranges $5 \leq N_{pts} \leq 50$ in steps of 5 and $1 \leq N_{Rx} \leq 9$ for odd numbers $N_{Rx}$ of transducers opposing Tx. For each ($N_{pts}$, $N_{Rx}$) combination, we retrain the model 10 times for each combination of training/testing folds, resulting in 10 RMSE values per ($N_{pts}$, $N_{Rx}$) combination. Figure 4(b) shows the mean (lines) and standard deviation (shaded) RMSE value as a function of $N_{pts}$ for several values of $N_{Rx}$. We observe that RMSE decreases from 17°C to 15°C on average, when the number $N_{Rx}$ of Rx increases from one to three. The decrease in RMSE is likely due to the fact that using measurements from additional Rx increases the amount of pertinent information provided to the CNN. The RMSE further decreases to 14°C by further increasing $N_{Rx} = 5$, for $N_{pts} \leq 10$, but other numbers of points result in an increase in RMSE from the models using the same $N_{pts}$ and $N_{Rx} = 1$ or 3. Subsequent increases to $N_{Rx} = 7$ and 9 were found to increase the RMSE on average to values of 24°C and 39°C, respectively. Here, increasing $N_{Rx}$ increases the available information at the cost of increasing the number of trainable CNN parameters. CNN training consists of using a gradient-based adaptive momentum (Adam) convex optimization algorithm. In general, the training process is a non-convex optimization problem, which means that the Adam solver will find a locally-optimal combination of CNN parameters, but it may not find the combination that is globally-optimal. Increasing the number of CNN parameters increases the dimensionality of the optimization problem, which decreases the likelihood that the globally-optimal set of parameters will be found. Thus, by increasing $N_{Rx}$ we balance the benefit of introducing additional information about the system with the increased dimensionality. For $N_{Rx} < 5$, the additional information is more beneficial than the increase in dimensionality, while for $N_{Rx} > 5$, the increase in dimensionality is more detrimental. Additionally, for $N_{Rx} > 5$, the additional information comes from transducers that are further from the opposing Rx. As a result, there is more interference between guided and bulk waves, and the amplitude of the first guided wave relatively high, as shown in Figure 1(c).

We note that the training and testing data was all measured or simulated on containers with nominally identical geometry and filled with nominally identical HE. The time required for the bursts to propagate between a Tx/Rx pair depends on the sound speed, which is temperature dependent, and the distance between the Tx/Rx. Thus, it is unlikely that the CNN, as presented in this manuscript, would be successful at estimating the temperature profile in a container with a significantly different shape or size. It may be possible to account for the size of the container by stretching/contracting the measured waveforms, but this is left for future work. Additionally, the dependence of the sound speed on temperature will differ between HE materials, which would introduce errors in the exact temperature values. However, most HE materials follow similar sound speed-temperature trends, i.e. decreasing sound speed with increasing temperature. Thus, it is likely that the CNN could measure the temperature profile trend, which could help identify if there was a solid-liquid transition (orange and yellow lines in Figure 4(a)), single phase with a temperature gradient (light blue line in Figure 4(a)), or constant temperature (red and dark blue lines in Figure 4(a)). Again, confirming and quantifying the performance of the CNN with different HE materials in training/testing is left for future work.

## 4. Conclusion

We present a novel technique for measuring internal temperature profiles by combining time-domain acoustic measurements and CNN processing. In contrast with existing temperature measurement methods our technique measures the interior temperature profile noninvasively, instead of requiring transducers to be placed inside or penetrate the container or being limited to measuring exterior surface temperatures. The technique is demonstrated on HE-filled containers as they are heated externally from ambient temperature until detonation/deflagration. Here, we introduce a hybrid measurement process, where we collect acoustic measurements experimentally, and measure the temperature profiles via finite element simulation. We then introduce a CNN that estimates the temperature at a specified number of points within the container based on the acoustic signals from one or more acoustic receiver. We observe that the CNN accurately estimates the temperatures, and it captures the temperature trends, which can provide critical information about phase, thermal gradient, etc. in the HE. Additionally, we find that increasing the number of receivers used measure the acoustic burst has competing effects of providing additional information about the temperature profile at the cost of increasing the model complexity. We observe the lowest RMSE of 15°C between the true and estimated temperatures by using three opposing receivers. In this study, training and testing data consisted of experiments and simulations on containers with nominally identical dimensions and materials. In the future, we will extend the data set to a range of dimensions and HE materials. We anticipate that this will require normalizing data to account for changes in shape. Additionally, it will increase the error in the absolute temperatures measured between different HE materials but will likely still provide crucial information such as whether or not there is a liquid-solid HE interface (is the HE partially melted?).

Thus, this work presents the first demonstration of using acoustics to measure internal thermal profiles in high-attenuation materials, through the material container. This technique has implications in a variety of applications including, assessing the safety status of HE materials, monitoring metal or plastic casting, determining the energy density in thermal storage capsules, and identifying abnormal battery operation, to name a few.

This work was supported by the U.S. Department of Energy through the Los Alamos National Laboratory. Los Alamos National Laboratory is operated by Triad National Security, LLC, for the National Nuclear Security Administration of U.S. Department of Energy (Contract No. 89233218CNA000001).